\documentclass{moriond}

\def\Journal#1#2#3#4{{#1} {\bf #2}, #3 (#4)}

\def\PLB{{\em Phys. Lett.}  B}
\def\PRL{\em Phys. Rev. Lett.}
\def\PRD{{\em Phys. Rev.} D}

\def\be{\begin{equation}}
\def\ee{\end{equation}}
\def\bea{\begin{eqnarray}}
\def\eea{\end{eqnarray}}

\newcommand{\Photo}{\includegraphics[height=35mm]{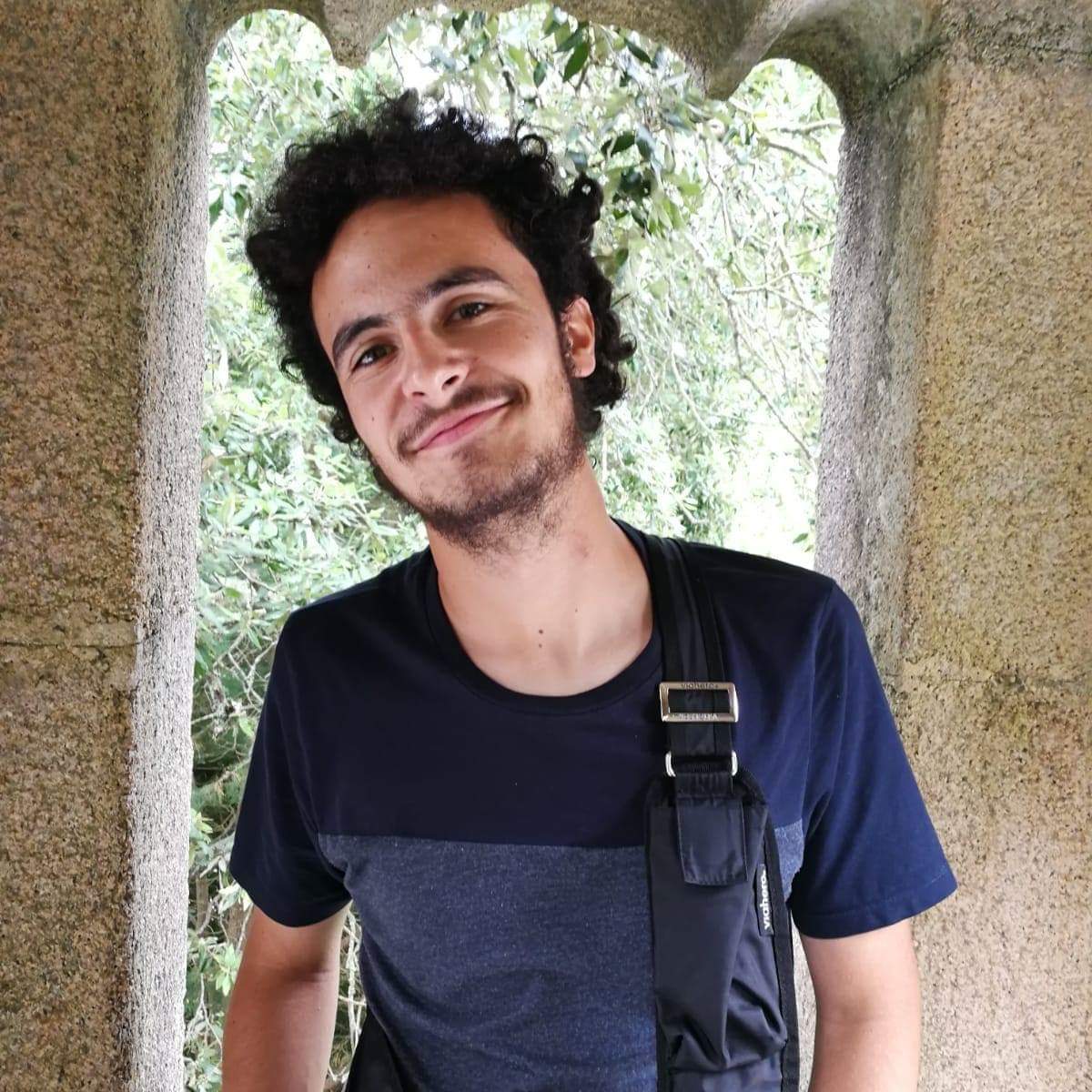}}

\begin{document}
\vspace*{4cm}
\title{GEV-SCALE NEUTRINOS: MESON INTERACTIONS AND DUNE SENSITIVITY}

\author{M. GONZÁLEZ-LÓPEZ}

\address{Departamento de Física Teórica and Instituto de Física Teórica UAM/CSIC,\\
Universidad Autónoma de Madrid, Cantoblanco, 28049, Madrid, Spain}
\maketitle\abstracts{The simplest extension of the SM to account for the observed neutrino masses and mixings is the addition of at least two singlet fermions (or right-handed neutrinos). If their masses lie at or below the GeV scale, such new fermions would be produced in meson decays. Similarly, provided they are sufficiently heavy, their decay channels may involve mesons in the final state.  In this work,  we provide consistent expressions for the relevant effective operators involving neutrinos and mesons with masses up to 2 GeV.  The associated effective Lagrangian has been implemented in FeynRules,  allowing for efficient simulation in event generators such as MadGraph5.  As an application of this setup, we numerically compute the expected sensitivity of the DUNE near detector to heavy neutrinos in the MeV-GeV range.
}

\section{Introduction}
The evidence of non-vanishing neutrino masses,  found in neutrino oscillations,  clearly requires an extension of the Standard Model (SM) in order to account for the experimentally measured masses and mixings.  The most natural way to generate neutrino masses consists on adding right-handed neutrinos (or heavy neutral leptons, HNLs) to the SM particle content.  In the minimal scenario, no extra interactions or gauge groups are introduced.  Thus,  HNLs are gauge singlets, and only communicate with the SM via mixing, replacing light neutrinos in any SM process in which the latter could be involved.  Neutrino flavor eigenstates will now also have a heavy component,  and the PMNS matrix will be enlarged, in order to include the elements which control the mixing between the heavy and light states (which is expected to be small).

As it is not necessarily connected to the electroweak scale, the mass of the right-handed neutrinos could lie in a very wide range,  from the eV to the GUT scale.  In the traditional type-I seesaw~\cite{mink,moha,yana,gell}, the measured light neutrino masses require very heavy HNLs (provided no fine-tuning is introduced in their Yukawa couplings).  Other versions of the seesaw mechanism, so called low-energy seesaws~\cite{moha2,moha3,berna,malin},  allow for HNLs close to the weak scale with no need for extremely small Yukawas.

Right-handed neutrinos in the GeV range are especially interesting.  They introduce no extra Higgs hierarchy problem, and can succesfully explain the baryon asymmetry of the universe through leptogenesis~\cite{pilar,asmaa,ghig}.  HNLs in this range are also relevant from a phenomenological point of view, as they can be searched for in different laboratory experiments, such as peak searches,  beam dumps or even colliders. 

In particular, near detectors of neutrino oscillations experiments could constitute an excellent framework to probe neutrinos in this mass scale.  Meson interactions are crucial in this scenario, as HNLs may be produced in heavy meson decays, and subsequently decay into visible states, including lighter mesons. With this motivation in mind, in this work we derive the low-energy effective operators controlling neutrino interactions with mesons, which, in this minimal extension of the SM,  only depend on the HNL masses and mixings. 

We will apply this setup to estimate the sensitivity of the Deep Underground Neutrino Experiment (DUNE) to HNLs.  This analysis will be performed in a 3 + 1 scenario, where only one heavy neutrino is introduced.  The measured neutrino masses and mixings require the inclusion of at least 2 HNL states; however, if we assume only one of them to mix sufficiently or to be light enough to be produced, the phenomenology is greatly simplifed, and reduces to the 3 + 1 case.

\section{Effective low-energy meson interactions}
In order to study neutrino production and decay via interactions with mesons,  it is necessary to compute the effective low-energy operators, once the weak bosons are integrated out and the corresponding hadronic matrix elements are properly parametrized.  This parametrization is done in terms of the meson decay constant and 4-momentum (for pseudoscalar mesons).

In order to obtain the effective operators, we first write down the relevant decay amplitudes, integrating out the corresponding weak boson and introducing Fermi's constant, $G_F$. Then, we insert the hadronic matrix element and translate the meson 4-momentum into a derivative acting on the lepton current. Finally,  we apply Dirac's equation, assuming that the charged lepton and the neutrino are on-shell.  Thus, we obtain Yukawa-like couplings controlled by the lepton and neutrino masses ($m_\alpha$ and $m_i$ respectively). For instance, the vertex involving a charged pion, a charged lepton and a neutrino mass eigenstate $n_i$ reads 
\begin{equation}
\mathcal{O}_{\pi \ell_\alpha \bar{n}_i } =  i \sqrt{2} G_F U_{\alpha i} V_{ud} f_\pi \bar \ell_\alpha  (m_\alpha P_L - m_i P_R) n_i  \pi^- + \textrm{h.c.} \, ,
\label{eq:effective-pion}
\end{equation}
where $U_{\alpha i}$ is the corresponding mixing matrix element and $f_\pi$ is the pion decay constant. This result can be generalized to other charged mesons, replacing the decay constant and the CKM element.  

We find qualitatively similar results for the couplings of neutral pseudoscalars to two neutrinos.  In the case of vector mesons, the procedure is similar, although the hadronic matrix elements are parametrized in terms of their polarization instead of their 4-momentum; thus, it is not possible to apply Dirac's equation on the lepton fields and the couplings are not proportional to the masses of the latter.

HNLs can also be produced in semileptonic meson decays.  In this case, the hadronic matrix element mediating the transition between the parent and the daughter mesons are parametrized in terms of two momentum dependent form factors. Although the procedure to obtain the effective operators is very similar to the one described above, the resulting vertices exhibit some differences. See Coloma \textit{et al.}~\cite{coloma} for details.

Apart from meson interactions, HNLs will also be involved in fully leptonic processes, controlled by the SM weak interactions via mixing, once the neutrino sector is extended.  This generalized weak Lagrangian and a full list of effective operators concerning HNLs and mesons can be found in Coloma \textit{et al.},~\cite{coloma} as well as expressions for the decay widths of processes controlled by these operators. We have implemented all these Feynman rules, for a 3 + 1 scenario, in a FeynRules~\cite{FR} model, which includes two versions,  describing Dirac and Majorana neutrinos respectively.
\section{DUNE Near Detector Sensitivity}
DUNE may constitute an excellent probe of HNLs in the GeV scale.  The protons impinging on the target will copiously produce charged mesons, which may decay into heavy neutrinos via mixing. Depending on their mass and the intensity of this mixing, these particles may travel hundreds of meters and decay inside the near detector (ND) into visible states, possibly containing lighter mesons, through a second mixing insertion. 

We will employ the effective theory derived in the previous section to estimate the sensitivity of the ND to the HNL mixing. We will do so by implementing the mentioned FeynRules model in MadGraph5\cite{mg5}, and generating events in which charged pseudoscalars decay into heavy neutrinos, as well as events in which the HNL decays into visible states (contanining lighter mesons or three leptons).  Regarding the parent meson fluxes, we obtain those of ligher parents (pions and kaons) from the GEANT4~\cite{geant} simulation developed by the DUNE collaboration, whereas for heavier parents ($D$, $D_s$ and $\tau$) we make use of both Pythia~\cite{pythia} and GEANT4.  We assume the geometry of the ND and the characteristics of the proton beam (an energy of 120 GeV and a rate of $1.1\cdot 10^{21}$ protons on target per year) specified in the DUNE Technical Design Report~\cite{dune}. 

\begin{figure}[htb!]
\centering
\includegraphics[height=5.742cm,keepaspectratio]{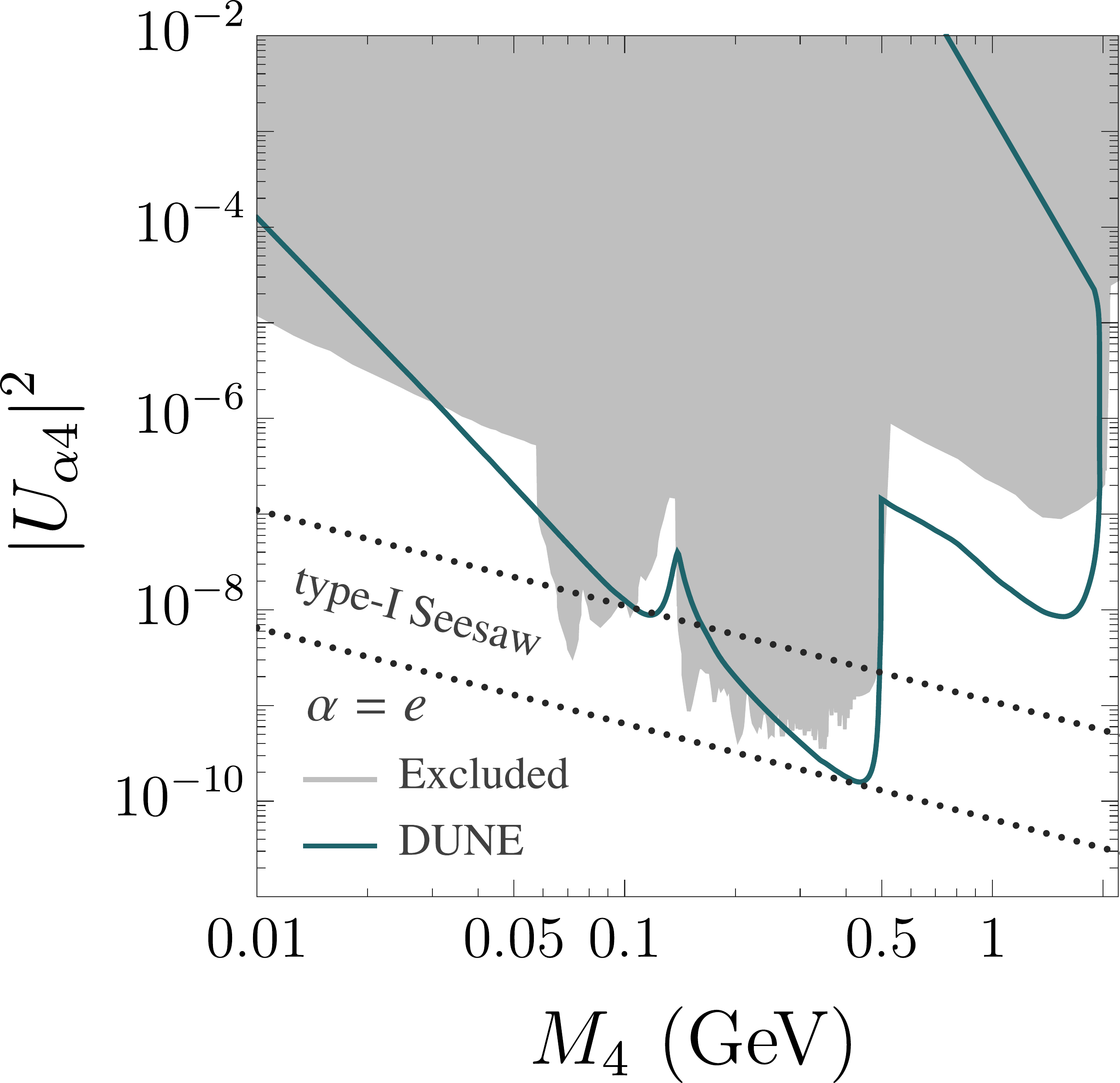} \hspace{-0.52 cm}
\includegraphics[height=5.56cm,keepaspectratio]{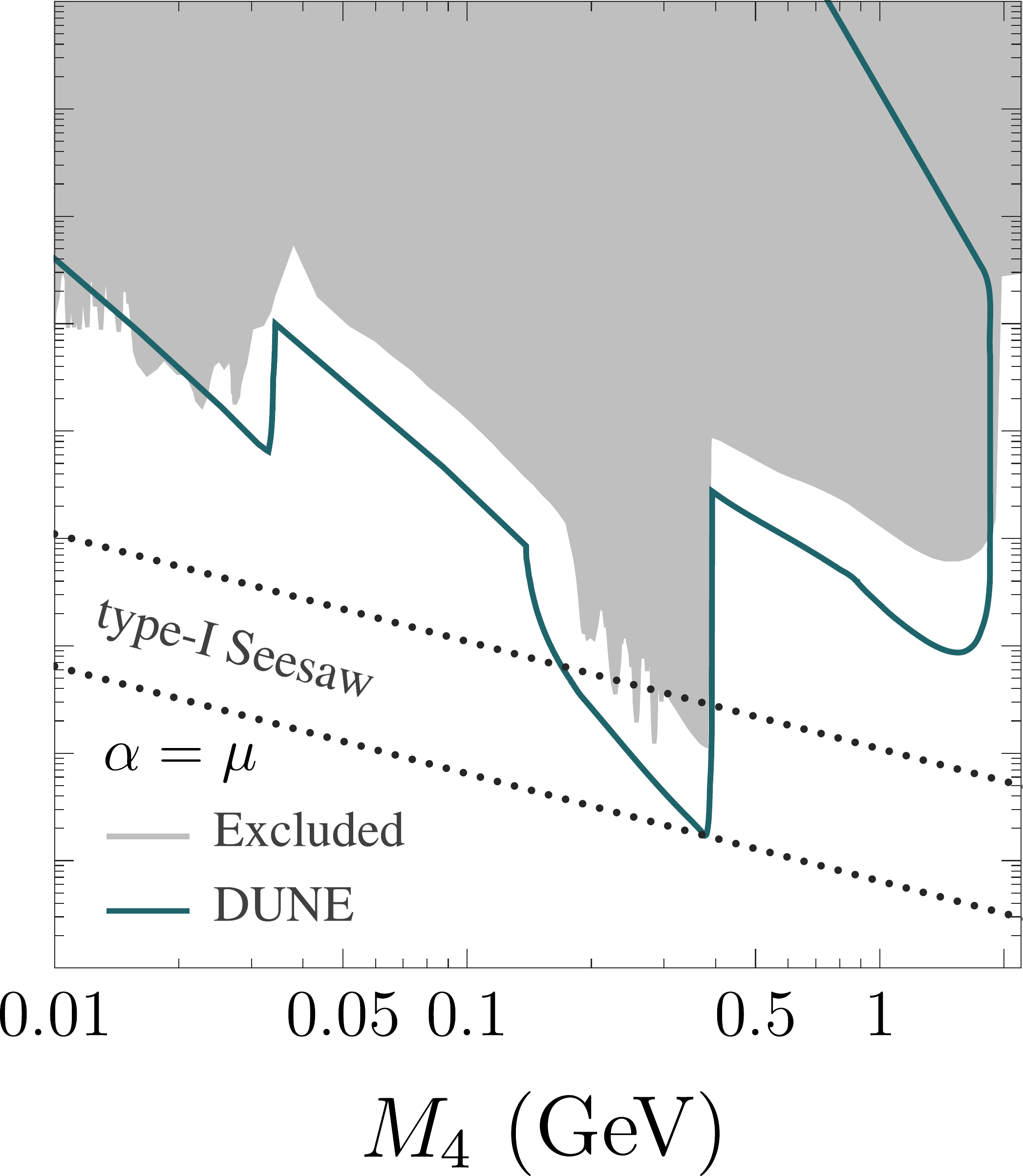} \hspace{-0.52 cm}
\includegraphics[height=5.56cm,keepaspectratio]{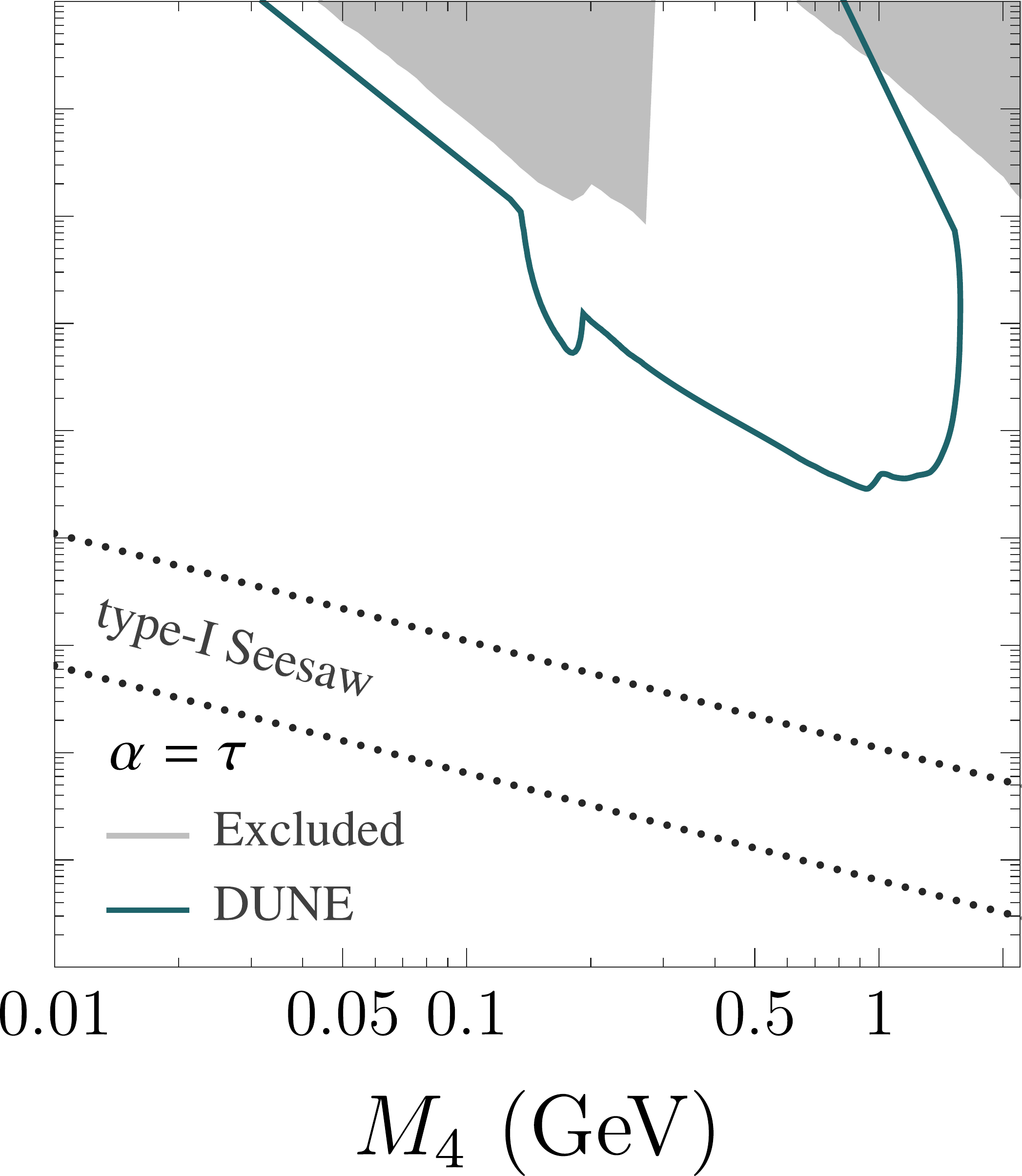}
\caption{DUNE ND sensitivity (at $90\%$ CL) to heavy neutrino mixings $\vert U_{\alpha 4}\vert^2$ as a function of its mass,  assuming mixing with only one flavor at a time and $7.7 \cdot 10^{21}$ protons on target.  Gray shaded areas are excluded by present experiments. The band limited by the dashed lines contains the predictions of the type-I seesaw mechanism.  The upper line is obtained applying the Katrin bound~\protect\cite{katrin} ($m_\nu\sim$ 1.1 eV), and the lower one setting $m_\nu=\sqrt{\Delta m^2_{atm}}\sim$ 0.05 eV.\label{fig:sens_total}}
\end{figure}

Figure~\ref{fig:sens_total} shows the 90$\%$ confidence level sensitivity of the DUNE ND to the HNL mixing as a function of its mass. We follow the Feldman and Cousins prescription~\cite{feld} for a Poisson distribution under the hypothesis of no observed events, which corresponds to no more than 2.44 HNL decay events taking place inside the detector (we assume the background to be negligible).  We consider a scenario in which the HNL only mixes simultaneously to one flavor, and display separately the sensitivity to $\vert U_{e4}\vert^2$, $\vert U_{\mu4}\vert^2$ and $\vert U_{\tau 4}\vert^2$.  Our results combine all possible HNL decay channels into visible states,  correspond to 7 years of data taking, and include a 20$\%$ signal efficiency, following Abe \textit{et al.}~\cite{t2k}.  We have computed the sensitivity to Dirac HNLs; in the Majorana case, it would improve by a factor 2, although the results would be qualitatively very similar.

We find that the DUNE ND will be able to probe new regions of the parameter space, which have not been accessed by previous experiments (gray shaded areas).  In fact,  current constraints could be improved in most of the considered mass range, even by several orders of magnitude in some regions. Interestingly,  it would possible to reach extremely small mixings, such as those predicted by the traditional type-I seesaw (dashed lines) in order to reproduce the measured neutrino masses. We have compared our results to those in similar studies, such as Ballett \textit{et al}.~\cite{silvia}, Berryman \textit{et al.}\cite{kevin} and others, finding overall agreement, with the exception of some details due to slight differences in the analyses (see Coloma \textit{et al.}~\cite{coloma} for a further discussion).

\section{Conclusions}
Heavy neutrinos in the GeV scale are particularly interesting, both from the theoretical and the experimental perspectives. On one side, these new particles may pose solutions to several open problems of the SM, such as neutrino masses or the BAU.  Besides,  neutrinos in this mass range could be probed in different types of laboratory experiments. In particular, they could be produced in meson decays, and decay visibly into SM particles, possibly lighter mesons. 

With that framework in mind,  we have derived the low-energy effective operators controlling the relevant interactions between HNLs and mesons, which, in the minimal setup, are only controlled by the heavy neutrino mixings and mass. We have implemented that effective Lagrangian in a FeynRules model and exported it to MadGraph5, allowing for efficient event generation. This way, we have computed the sensitivity of the DUNE ND to the heavy neutrino mixing as a function of its mass. We have found that this facility could greatly improve the current bounds,  being sensitive to extremely small mixings, such as those predicted by the type-I seesaw.

\section*{Acknowledgments}
I would like to thank the co-authors of this work,  Pilar Coloma, Enrique Fernández-Martínez, Josu Hernández-García and Zarko Pavlovic, it was a pleasure to write this paper with you.  I am also grateful to the organizers of the 55th Rencontres de Moriond for the kind invitation.

\end{document}